\documentclass[modern]{lsstdescnote}
\usepackage{lsstdesc_macros}
\usepackage{appendix}
\usepackage{graphicx}
\usepackage{hyperref}
\usepackage{tocbasic}

\pdfoutput=1 
\usepackage[utf8]{inputenc}
\DeclareUnicodeCharacter{2212}{-}

\DeclareTOCStyleEntry[beforeskip=-16pt, entryformat={},
                     linefill=\TOCLineLeaderFill]{tocline}{section}
\DeclareTOCStyleEntry[beforeskip=-16pt]{tocline}{subsection}
\usepackage{booktabs}
\usepackage[most]{tcolorbox}
\tcbset{colframe=black,
        size=small,
        arc=0pt,
        boxrule=0.5pt}
\hypersetup{
    colorlinks,
    citecolor=black,
    filecolor=black,
    linkcolor=black,
    urlcolor=black
}

\newcommand{\votelss}{\begin{tcolorbox}[colback=yellow!40!white,hbox,nobeforeafter]LSS\end{tcolorbox}}
\newcommand{\votewl}{\begin{tcolorbox}[colback=red!40!white,hbox,nobeforeafter]WL\end{tcolorbox}}
\newcommand{\votepz}{\begin{tcolorbox}[colback=blue!40!white,hbox,nobeforeafter]PZ\end{tcolorbox}}
\newcommand{\votesn}{\begin{tcolorbox}[colback=orange!40!white,hbox,nobeforeafter]SN\end{tcolorbox}}
\newcommand{\votesl}{\begin{tcolorbox}[colback=PineGreen!40!white,hbox,nobeforeafter]SL\end{tcolorbox}}
\newcommand{\votecl}{\begin{tcolorbox}[colback=green!40!white,hbox,nobeforeafter]CL\end{tcolorbox}}

\newcommand{\votedkm}{\begin{tcolorbox}[colback=pink!40!white,hbox,nobeforeafter]DKM\end{tcolorbox}}
\newcommand{\votepc}{\begin{tcolorbox}[colback=cyan!40!white,hbox,nobeforeafter]PC\end{tcolorbox}}

\graphicspath{{./figures/}}

\begin{document}

\title{Recommended Target Fields for Commissioning the Vera C. Rubin Observatory}
\author[0000-0002-6445-0559]{A. Amon}
\affiliation{Kavli Institute for Particle Astrophysics and Cosmology and Department of Physics, Stanford University, Stanford, CA 94305, USA}

\author[0000-0001-8156-0429]{K. Bechtol}
\affiliation{Department of Physics, University of Wisconsin-Madison, Madison, WI 53706, USA}

\author[0000-0001-5576-8189]{A.J. Connolly}
\affiliation{DIRAC Institute and Department of Astronomy, University of Washington, Seattle WA 98195, USA}

\author[0000-0002-5296-4720]{S.W. Digel}
\affiliation{SLAC National Accelerator Laboratory, 2575 Sand Hill Road, Menlo Park, CA 94025, USA}

\author[0000-0001-8251-933X]{A.~Drlica-Wagner}
\affiliation{Fermi National Accelerator Laboratory, P.O.~Box 500, Batavia, IL 60510, USA}
\affiliation{University of Chicago, Department of Astronomy and Astrophysics, University of Chicago, Chicago, IL 60637, USA}

\author[0000-0003-1530-8713]{E.~Gawiser}
\affiliation{Department of Physics and Astronomy, Rutgers the State University of New Jersey, 136 Frelinghuysen Road, Piscataway, NJ 08854, USA}

\author[0000-0002-4179-5175]{M.~Jarvis}
\affiliation{Department of Physics and Astronomy, University of Pennsylvania, Philadelphia, PA 19104, USA}

\author[0000-0001-8738-6011]{S.W.~Jha}
\affiliation{Department of Physics and Astronomy, Rutgers the State University of New Jersey, 136 Frelinghuysen Road, Piscataway, NJ 08854, USA}

\author[0000-0002-3881-7724]{A. von der Linden}
\affiliation{Department of Physics and Astronomy, Stony Brook University, Stony Brook, NY 11794, USA}

\author[0000-0001-8716-6561]{M. Moniez}
\affiliation{IJCLab, Bat. 100, 15 rue Georges Clemenceau, 91405 Orsay Cedex, France}

\author[0000-0001-6022-0484]{G. Narayan}
\affiliation{Department of Astronomy, University of Illinois at Urbana-Champaign, Urbana, IL 61801, USA}
\affiliation{Center for Astrophysical Surveys, National Center for Supercomputing Applications, Urbana, IL 61801, USA}

\author[0000-0001-7029-7901]{N. Regnault}
\affiliation{Laboratoire de Physique Nucleaire et de Hautes Energies (LPNHE), CNRS/IN2P3, Sorbonne Universit\'e et Universit\'e Paris-Diderot,
4 place Jussieu, 75005 Paris}

\author[0000-0002-1831-1953]{I. Sevilla-Noarbe}
\affiliation{Centro de Investigaciones Energ\'eticas, Medioambientales y Tecnol\'ogicas (CIEMAT), Madrid, Spain}

\author[0000-0002-5091-0470]{S.J.~Schmidt}
\affiliation{Department of Physics, University of California, Davis CA 95616, USA}

\author[0000-0001-5568-6052]{S.~H.~Suyu}
\affiliation{Max-Planck-Institut f{\"u}r Astrophysik, Karl-Schwarzschild-Str.~1, 85748 Garching, Germany}
\affiliation{Physik-Department, Technische Universit\"at M\"unchen, James-Franck-Stra\ss{}e~1, 85748 Garching, Germany}
\affiliation{Academia Sinica Institute of Astronomy and Astrophysics (ASIAA), 11F of ASMAB, No.1, Section 4, Roosevelt Road, Taipei 10617, Taiwan}

\author[0000-0003-2035-2380]{C.W. Walter}
\affiliation{Department of Physics, Duke University, Durham NC 27708, USA}
\collaboration{(The Dark Energy Science Collaboration)}
\date{\today}

\begin{abstract}
  
The commissioning team for the Vera C. Rubin observatory is planning a set of engineering and science verification observations with the Legacy Survey of Space and Time (LSST) commissioning camera~\citep{2018SPIE10700E..3DH} and then the Rubin Observatory LSST Camera~\citep{2018SPIE10705E..0DR}.  The time frame for these observations is not yet fixed, and the commissioning team will have flexibility in selecting fields to observe.  In this document, the Dark Energy Science Collaboration (DESC) Commissioning Working Group presents a prioritized list of target fields appropriate for testing various aspects of DESC-relevant science performance, grouped by season for visibility from Rubin Observatory at Cerro Pachon. Our recommended fields include Deep-Drilling fields (DDFs) to full LSST depth for photo-$z$ and shape calibration purposes, HST imaging fields to full depth for deblending studies, and an $\sim$200 square degree area to 1-year depth in several filters for higher-level validation of wide-area science cases for DESC. We also anticipate that  commissioning observations will be needed for template building for transient science over a broad RA range. We include detailed descriptions of our recommended fields along with associated references.  We are optimistic that this document will continue to be useful during LSST operations, as it provides a comprehensive list of overlapping data-sets and the references describing them.
  
\end{abstract}

\maketitle

\tableofcontents

\clearpage

\section{Summary of DESC recommendations}

The Vera C. Rubin Observatory will conduct a 10-year Legacy Survey of Space and Time \citep[LSST,\footnote{\href{http://www.lsst.org}{http://www.lsst.org}}][]{2019ApJ...873..111I} to advance multiple frontiers of astrophysics and cosmology research. The Rubin Observatory is currently under construction, and the transition to LSST operations is expected to begin at the end of 2022. Before and during that time, both the hardware and software systems need verification that they have been built to specification, and validation that the entire system can be used to achieve its science objectives. 
As part of this commissioning effort, the Rubin Observatory Project has planned a series of on-sky observing campaigns.
The exact field locations and details of the commissioning survey strategy (e.g., cadence, integrated exposure, coverage area, and band selection) have not been fully defined and will be under consideration even during commissioning.

The Rubin Observatory Project has informally solicited input regarding scientific optimization of the on-sky observing strategy during commissioning\footnote{\href{https://project.lsst.org/meetings/lsst2019/content/community-project-coordination-science-verification-during-commissioning}{See, for example, https://project.lsst.org/meetings/lsst2019/content/community-project-coordination-science-verification-during-commissioning}}. It has indicated that guidance from the science community would be taken into consideration to enhance commissioning science validation, alongside both constraints from the overall cost and schedule of the construction project and the need to technically verify and optimize the as-built system.
We the Dark Energy Science Collaboration (DESC\footnote{\href{http://www.lsstdesc.org}{http://www.lsstdesc.org}}), one of eight LSST Science Collaborations, have a particular focus on studies of dark energy along with dark matter, neutrino properties, inflation, and other topics in fundamental physics.
In this note, we identify a set of candidate target fields and provide resources (see Appendix~\ref{sec:Appendix-Accompanying-Material}) that may be useful to the Rubin Observatory Project, and to other LSST Science Collaborations, to facilitate science validation activities with commissioning data. 
Our prioritized high-level list of recommendations for the commissioning observing strategy is summarized below.  Details are provided in the sections that follow.

\begin{enumerate}
\item Observe an LSST Deep Drilling Field (DDF) to at least 10-year equivalent depth of the LSST Wide-Fast-Deep (WFD) survey in multiple bands. The DDFs overlap existing external datasets and are advantageous for many extragalactic studies. 
\item Start building templates for supernova science (and other transients). Start with a DDF and then expand across as wide a range of right ascension as possible.  Repeated observations in the same filters spaced by 1 to 4 nights will allow science validation with supernova lightcurves. 
\item Observe contiguous wide-field areas ($\gtrsim100$~deg$^2$) in equatorial regions such as SDSS Stripe 82 in multiple bands (e.g., $gri[z]$) to 1-year LSST WFD equivalent depth.
\item Observe fields with deep high-resolution optical space-based imaging (e.g., HST) to test object detection and deblending, galaxy shape measurement, and star-galaxy separation.
\item On photometric nights, alternate multiband imaging of DDFs and spectrophotometric standards to validate and enhance photometric calibration.
\end{enumerate}

We hope that this document and associated tools will also help to engage the other LSST Science Collaborations in the process of planning for commissioning.  We anticipate that this note may be complemented by similar documents from other Science Collaborations, and that the proposed fields for observation will be integrated into a common framework that reflects the broad LSST science community.

\section{Overview of planned commissioning observations}
\label{sec:overview}
We here summarize our understanding of Rubin Observatory Project plans for system integration and commissioning \citep[][]{LSE-79}.  The plans include tests to verify, validate, and characterize the technical and scientific performance of the observatory.\footnote{Details of the Rubin Observatory system-level specifications are found in the LSST System Requirements (\href{http://ls.st/LSE-29}{LSE-29}) and Observatory System Specifications (\href{http://ls.st/LSE-30}{LSE-30}) documents.}
The commissioning plan includes three phases of increasing on-sky observing capability:
(1) system integration and testing with the commissioning camera, ``ComCam'',
(2) system integration and testing with the full LSST Camera, ``LSSTCam'', and
(3) two Science Verification Surveys in which the observatory operates in a continuous mode similar to early LSST operations.
ComCam uses a single raft of 9 4K $\times$ 4K Science CCDs, and replicates most of the interfaces and functionality of the full 21-raft LSSTCam \citep{2018SPIE10700E..3DH}.
The ComCam field of view is 0.7~deg $\times$ 0.7~deg, compared to the 3~deg diameter field of view of the LSSTCam.
The ComCam filter changer holds up to three filters at a time, relative to five filters for LSSTCam.

The two system integration and testing phases with ComCam and LSSTCam, respectively, are scheduled to include opportunistic science imaging interspersed with engineering activities.
Potential science observations include star flats, as well as deep imaging (several hundred visits per band) of selected fields to investigate systematic effects at full LSST survey depth.

The two Science Verification Surveys are designed to produce and evaluate the Prompt data products and the Data Release data products, respectively. 
The first Science Verification Survey will cover a region of sky approaching 10\% of the LSST WFD survey area ($\gtrsim1000$~deg$^2$) with an integrated exposure in several bands similar to that of the first two years of the WFD survey ($\sim30$ visits per band). 
The region is planned to be observed in two phases: first for template generation and then returning to the same region to test difference imaging and alert generation.
The second Science Verification Survey is planned to cover a region of sky roughly 1\% of the LSST WFD survey area to nominal 10-year equivalent exposure in multiple bands.
The exact schedule and duration of each of these phases will be determined by the needs of the Project, and thus the area, depth, and band coverage may be adjusted from the baseline plan. 

A flexible observing plan is needed given the uncertainty around the commissioning schedule.
In the following sections, we identify fields that would be useful to observe under various instrumental configurations and environmental conditions, and organize the candidate fields according to season and appropriateness for particular tests. 
Our field suggestions could be usable both in planning multi-night observing campaigns, and in situations where the Project commissioning team finds itself with a few hours for opportunistic science observations during engineering time.

\section{DESC perspectives on candidate field selection}

\emph{During commissioning, the primary interests of the DESC are to test the robustness of our analysis pipelines and to search for and characterize systematic effects/anomalies in the data.} We do not expect to produce competitive cosmological results using commissioning data.

By the time of on-sky observations during commissioning, we will have tested many of the DESC analysis pipelines on precursor datasets from other surveys (e.g., HSC, DES, KiDS), as well as a dedicated suite of pixel-level LSST image simulations such as DESC Simulation Data Challenges.
During commissioning, we will evaluate DESC analysis pipelines with Rubin Observatory data, evaluate performance with respect to DESC science goals \citep{2018arXiv180901669T}, and could potentially provide feedback to the Rubin Observatory project that would help to optimize hardware and software performance during LSST operations.
The DESC has already committed effort to perform science validation studies as commissioning data become available.\footnote{DESC Science Roadmap: \url{https://lsstdesc.org/assets/pdf/docs/DESC_SRM_latest.pdf}}
Many tests that the DESC plans could be undertaken on fields chosen for other purposes by the Rubin Observatory commissioning team.  For DESC science verification and validation specifically, the most important considerations are that the fields overlap with regions of the sky that are well measured by other surveys in multiple wavelengths, have existing spectroscopy, and/or have space-based imaging. 

The field recommendations in this document arise from discussions within DESC. Details pertaining to different scientific cases are provided in Appendix~\ref{sec:Appendix-Science-Drivers}.  We have identified several topics of primary importance to address in the Rubin Observatory commissioning phase:  

\begin{itemize}
\item Photo-$z$ training and calibration;
\item PSF characterization and galaxy shape calibration for weak lensing shear measurements;
\item Testing alert production pipelines and community broker interactions;
\item Production of template images for difference imaging that would enable realtime alert generation in the first year of LSST Operations (LOY1);
\item Evaluating photometric repeatability to assess achievability of a millimag goal;
 \item Understanding selection effects affecting object detection using well-curated external datasets extending to similar or greater depths;
\item Studying the impact of deblending on different science probes and products (e.g., single-object detection for wide-field probes, photo-$z$ biases, shear biases);
\item Accuracy of star-galaxy classifications;
\item Observatory telemetry and other metadata needed to construct `survey property maps' useful for large-scale statistical analyses.
\end{itemize}

As noted in Section~\ref{sec:overview}, due to the finite period of commissioning, templates for difference imaging will be obtained for only a subset of the full LSST survey area. We emphasize that obtaining observations to make templates for the DDFs during commissioning are crucial for DESC Supernova Working Group (SN WG) science operations in LOY1.

For the studies above, several selection requirements were considered when choosing fields to recommend for observation:

\begin{itemize}
    \item Observability from Cerro Pachon;
    \item Distance to the Galactic plane (we would ideally sample a range of Galactic latitudes to study effects of high stellar density and interstellar extinction); 
    \item Availability of previous survey data, spectroscopic information, spectrophotometric standards, deep multi-band imaging, and/or deep space-based imaging, and 
    \item Presence of rare astrophysical objects, such as strong lenses or rich galaxy clusters;
    \item Existence of an LSST DDF.
\end{itemize}

Several external datasets, such as those from GALEX, Gaia, 2MASS, and WISE, have all-sky coverage and thus require no special observing strategy to achieve overlap.

If possible, it may be useful to select fields that are visible over several months of the commissioning period in order to have a longer time baseline to characterize transient, variable, and moving objects. 
For time-domain studies, it is preferable to have smaller number of fields that are repeatedly observed in a few filters, rather than an assortment without regularly cadenced observations.

\section{Field locations and observability}
\label{sec:observability}

To enable the goals outlined above, each DESC WG has identified a set of target fields visible from Cerro Pachon that contain precursor data sets or otherwise will allow them to test their analysis pipelines. The information on all fields has been collated into a multi-tabbed spreadsheet which is programmatically used to generate tables and observing strategies. The spreadsheets and Jupyter notebooks used to produce the tables and figures in this document are also available to the reader. Please see Appendix~\ref{sec:Appendix-Accompanying-Material} for more details. We have collated and merged each group's requests so as to understand if more than one group has requested the same field. This information was used to assist us in prioritization.  Detailed information from each group and every field can be found in the relevant tab on the spreadsheet.  Additionally, we have collated an alphabetical listing of fields with information including references for each of them that can be found in Appendix~\ref{sec:Appendix-field-list}.

\begin{figure}[!htb]
  \begin{center}
   \includegraphics[width=5.5in]{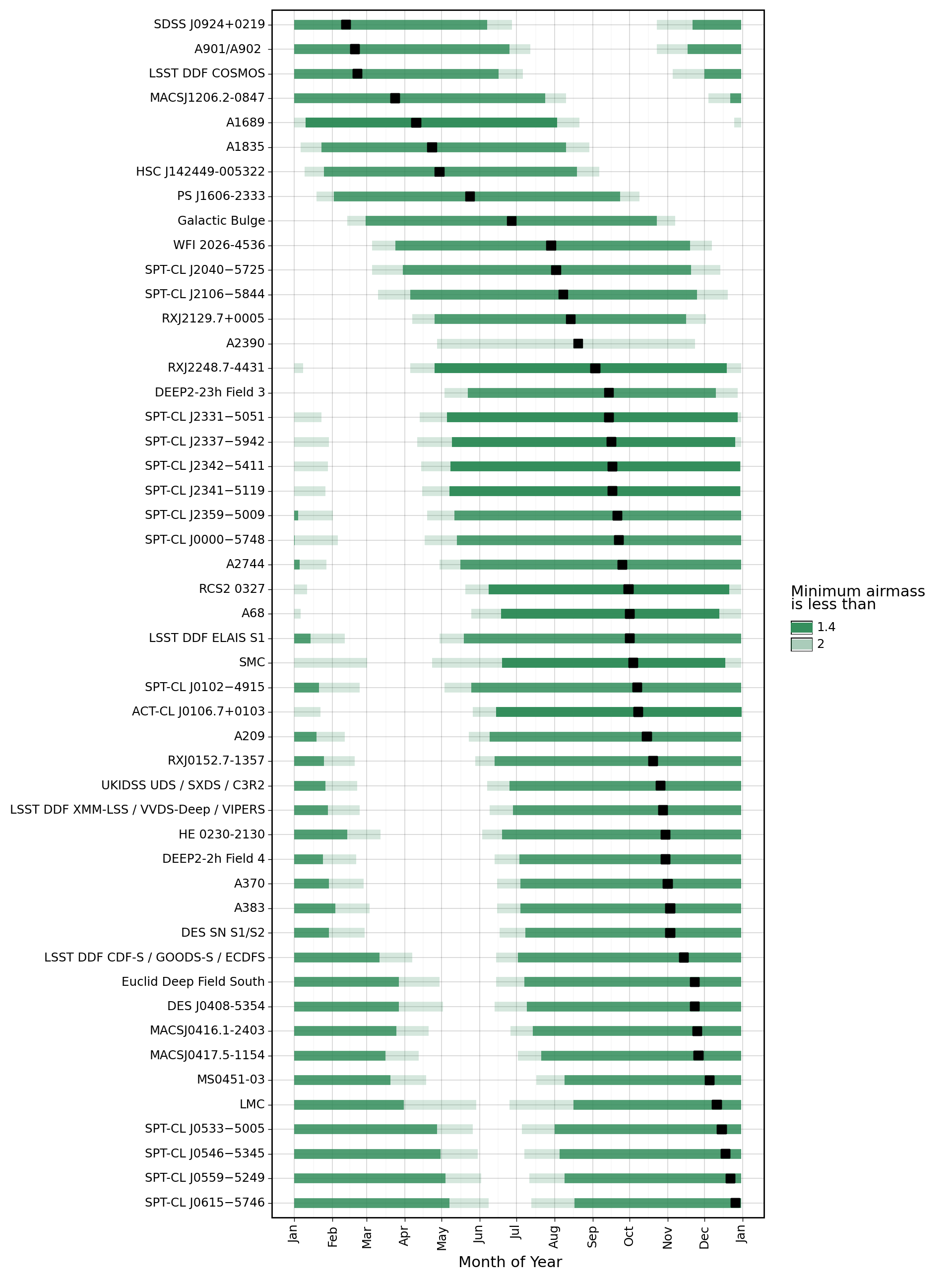}
    \caption{Visibility as a function of month for targets in Table~\ref{tab:observability_date} sorted by optimal yearly observing date. For each target, green and light green bands show when the target has air-mass less than 1.4 (solid) or 2.0 (light) at the time of optimal observing. The optimal observing date for each target is indicated with a black square.}
    \label{fig:visibility}
  \end{center}
\end{figure}

In Table~\ref{tab:observability_date}, for each of the small-area candidate fields, we show the airmass at the date of optimal observability along with a range of dates for which the airmass is small enough for good seeing. The table is ordered by the time of year for optimal observation. Figure~\ref{fig:visibility} graphically represents the same information and compactly summarizes when each fields is visible from Cerro Pachon.


\begin{table}[htbp]
\tiny
\begin{tabular}{lrrcccccc}
\toprule
        Field Name & RA (J2000) & Dec (J2000) & Optimal Date & Optimal Airmass & $T_{1.4}$ & $T_{2.0}$ & Dates $\sec(z) < 1.4$ & Dates $\sec(z) < 2.0$ \\
       ($\dagger$ indicates LSSTCam FOV) & (deg) & (deg) & Date (DOY) & & (hr) & (hr) & & \\
\midrule

                                   SDSS J0924+0219 & 141.23 & 2.32 & Feb-12 (043) & 1.19 & 4.1 & 7.0 & Nov-21 to Jun-07 & Oct-23 to Jun-27 \\
                                        A901/A902  & 149.10 & -10.00 & Feb-19 (050) & 1.07 & 5.6 & 7.9 & Nov-17 to Jun-25 & Oct-23 to Jul-12 \\
                         LSST DDF COSMOS$^\dagger$ & 150.12 & 2.21 & Feb-21 (052) & 1.18 & 4.2 & 7.0 & Dec-01 to Jun-16 & Nov-05 to Jul-06 \\
                                  MACSJ1206.2-0847 & 181.55 & -8.80 & Mar-24 (083) & 1.07 & 5.5 & 8.0 & Dec-22 to Jul-24 & Dec-04 to Aug-10 \\
                                             A1689 & 197.88 & -1.34 & Apr-10 (100) & 1.14 & 4.7 & 7.3 & Jan-10 to Aug-03 & Dec-25 to Aug-21 \\
                                             A1835 & 210.26 & 2.88 & Apr-23 (113) & 1.19 & 4.0 & 6.9 & Jan-23 to Aug-10 & Jan-06 to Aug-29 \\
                                HSC J142449-005322 & 216.20 & -0.89 & Apr-29 (119) & 1.15 & 4.6 & 7.2 & Jan-25 to Aug-19 & Jan-09 to Sep-06 \\
                                     PS J1606-2333 & 241.50 & -23.56 & May-24 (144) & 1.01 & 6.5 & 8.9 & Feb-02 to Sep-23 & Jan-19 to Oct-09 \\
                          Galactic Bulge$^\dagger$ & 275.00 & -27.50 & Jun-27 (178) & 1.00 & 6.6 & 9.2 & Feb-28 to Oct-23 & Feb-13 to Nov-07 \\
                                     WFI 2026-4536 & 306.54 & -45.61 & Jul-29 (210) & 1.04 & 7.1 & 10.1 & Mar-24 to Nov-19 & Mar-05 to Dec-07 \\
                                 SPT-CL J2040−5725 & 310.06 & -57.43 & Aug-02 (214) & 1.12 & 6.7 & 10.4 & Mar-30 to Nov-20 & Mar-05 to Dec-14 \\
                                 SPT-CL J2106−5844 & 316.52 & -58.75 & Aug-08 (220) & 1.14 & 6.6 & 10.4 & Apr-05 to Nov-25 & Mar-10 to Dec-20 \\
                                    RXJ2129.7+0005 & 322.42 & 0.09 & Aug-14 (226) & 1.16 & 4.4 & 7.1 & Apr-25 to Nov-16 & Apr-07 to Dec-02 \\
                                             A2390 & 328.40 & 17.70 & Aug-20 (232) & 1.50 & 0.0 & 4.8 & None & Apr-27 to Nov-23 \\
                                    RXJ2248.7-4431 & 342.18 & -44.53 & Sep-03 (246) & 1.03 & 7.1 & 9.8 & Apr-25 to Dec-19 & Apr-05 to Jan-08 \\
                       DEEP2-23h Field 3$^\dagger$ & 352.50 & 0.00 & Sep-14 (257) & 1.16 & 4.4 & 7.2 & May-22 to Dec-10 & May-03 to Dec-28 \\
                                 SPT-CL J2331−5051 & 352.96 & -50.86 & Sep-14 (257) & 1.07 & 7.0 & 9.5 & May-05 to Dec-28 & Apr-13 to Jan-23 \\
                                 SPT-CL J2337−5942 & 354.35 & -59.70 & Sep-16 (259) & 1.15 & 6.6 & 9.4 & May-09 to Dec-26 & Apr-11 to Jan-29 \\
                                 SPT-CL J2341−5119 & 355.30 & -51.33 & Sep-17 (260) & 1.07 & 7.0 & 9.4 & May-07 to Dec-30 & Apr-15 to Jan-26 \\
                                 SPT-CL J2342−5411 & 355.69 & -54.19 & Sep-17 (260) & 1.09 & 6.9 & 9.4 & May-08 to Dec-30 & Apr-14 to Jan-28 \\
                                 SPT-CL J2359−5009 & 359.92 & -50.16 & Sep-21 (264) & 1.06 & 7.1 & 9.3 & May-11 to Jan-04 & Apr-19 to Feb-01 \\
                                 SPT-CL J0000−5748 & 0.25 & -57.81 & Sep-22 (265) & 1.13 & 6.7 & 9.2 & May-13 to Jan-01 & Apr-17 to Feb-05 \\
                                             A2744 & 3.59 & -30.40 & Sep-25 (268) & 1.00 & 6.8 & 9.0 & May-16 to Jan-05 & Apr-29 to Jan-27 \\
                                         RCS2 0327 & 8.41 & 2.72 & Sep-30 (273) & 1.19 & 4.1 & 6.9 & Jun-08 to Dec-21 & May-20 to Jan-11 \\
                       LSST DDF ELAIS S1$^\dagger$ & 9.45 & -44.00 & Oct-01 (274) & 1.03 & 7.1 & 8.9 & May-19 to Jan-14 & Apr-29 to Feb-11 \\
                                               A68 & 9.28 & 9.16 & Oct-01 (274) & 1.30 & 2.7 & 6.1 & Jun-18 to Dec-13 & May-25 to Jan-06 \\
                                     SMC$^\dagger$ & 12.50 & -73.00 & Oct-04 (277) & 1.36 & 3.0 & 8.8 & Jun-19 to Dec-18 & Apr-23 to Mar-01 \\
                                 SPT-CL J0102−4915 & 15.73 & -49.26 & Oct-07 (280) & 1.06 & 7.1 & 8.7 & May-25 to Jan-21 & May-03 to Feb-23 \\
                               ACT-CL J0106.7+0103 & 16.71 & 1.07 & Oct-08 (281) & 1.17 & 4.3 & 7.0 & Jun-14 to Dec-31 & May-26 to Jan-22 \\
                                              A209 & 22.97 & -13.61 & Oct-15 (288) & 1.04 & 5.8 & 8.1 & Jun-09 to Jan-19 & May-23 to Feb-11 \\
                                    RXJ0152.7-1357 & 28.17 & -13.96 & Oct-20 (293) & 1.04 & 5.9 & 8.0 & Jun-13 to Jan-25 & May-28 to Feb-19 \\
                UKIDSS UDS / SXDS / C3R2$^\dagger$ & 34.50 & -5.00 & Oct-26 (299) & 1.11 & 5.0 & 7.6 & Jun-25 to Jan-26 & Jun-07 to Feb-21 \\
   LSST DDF XMM-LSS / VVDS-Deep / VIPERS$^\dagger$ & 36.45 & -4.60 & Oct-28 (301) & 1.11 & 5.0 & 7.5 & Jun-28 to Jan-28 & Jun-09 to Feb-23 \\
                                      HE 0230-2130 & 38.14 & -21.29 & Oct-30 (303) & 1.01 & 6.4 & 7.9 & Jun-19 to Feb-13 & Jun-03 to Mar-12 \\
                        DEEP2-2h Field 4$^\dagger$ & 37.50 & 0.00 & Oct-30 (303) & 1.16 & 4.4 & 7.1 & Jul-03 to Jan-24 & Jun-13 to Feb-20 \\
                                              A370 & 39.97 & -1.57 & Nov-01 (305) & 1.14 & 4.7 & 7.3 & Jul-04 to Jan-29 & Jun-15 to Feb-26 \\
                                              A383 & 42.01 & -3.53 & Nov-03 (307) & 1.12 & 4.9 & 7.4 & Jul-04 to Feb-03 & Jun-15 to Mar-03 \\
                            DES SN S1/S2$^\dagger$ & 42.00 & 0.00 & Nov-03 (307) & 1.16 & 4.5 & 7.2 & Jul-08 to Jan-29 & Jun-17 to Feb-27 \\
        LSST DDF CDF-S / GOODS-S / ECDFS$^\dagger$ & 53.12 & -27.81 & Nov-14 (318) & 1.00 & 6.7 & 7.4 & Jul-02 to Mar-11 & Jun-14 to Apr-07 \\
                 Euclid Deep Field South$^\dagger$ & 61.24 & -48.42 & Nov-23 (327) & 1.05 & 6.9 & 7.1 & Jul-07 to Mar-27 & Jun-14 to Apr-29 \\
                                    DES J0408-5354 & 62.09 & -53.90 & Nov-23 (327) & 1.09 & 6.8 & 7.1 & Jul-09 to Mar-27 & Jun-13 to May-02 \\
                                  MACSJ0416.1-2403 & 64.04 & -24.07 & Nov-25 (329) & 1.01 & 6.6 & 7.1 & Jul-14 to Mar-25 & Jun-26 to Apr-20 \\
                                  MACSJ0417.5-1154 & 64.38 & -11.78 & Nov-26 (330) & 1.05 & 5.8 & 7.0 & Jul-21 to Mar-16 & Jul-02 to Apr-12 \\
                                         MS0451-03 & 73.55 & -3.01 & Dec-05 (339) & 1.12 & 4.8 & 6.8 & Aug-09 to Mar-20 & Jul-17 to Apr-18 \\
                                     LMC$^\dagger$ & 80.00 & -69.00 & Dec-11 (345) & 1.28 & 5.0 & 6.7 & Aug-16 to Mar-31 & Jun-25 to May-29 \\
                                 SPT-CL J0533−5005 & 83.40 & -50.09 & Dec-15 (349) & 1.06 & 6.7 & 6.7 & Aug-01 to Apr-27 & Jul-05 to May-26 \\
                                 SPT-CL J0546−5345 & 86.65 & -53.76 & Dec-18 (352) & 1.09 & 6.7 & 6.7 & Aug-05 to Apr-30 & Jul-07 to May-30 \\
                                 SPT-CL J0559−5249 & 89.93 & -52.83 & Dec-22 (356) & 1.08 & 6.7 & 6.7 & Aug-09 to May-04 & Jul-11 to Jun-02 \\
                                 SPT-CL J0615−5746 & 93.97 & -57.78 & Dec-26 (360) & 1.13 & 6.7 & 6.7 & Aug-17 to May-07 & Jul-13 to Jun-08 \\
\bottomrule
\end{tabular}
\caption{Field observability (ordered by optimal date) throughout the calendar year. $T_{1.4}$ ($T_{2.0}$) represents 
the time in hours that a given field is observable at an airmass less than 1.4 (2.0) on the optimal night of the year. 
The two rightmost columns provide the range of calendar dates for which the field is observable at some point during the night 
(solar elevation angle $< -18^{\circ}$) at an airmass $\sec(z)$ below the indicated threshold. A $\dagger$ symbol beside the field name 
indicates that the field is optimally observed with the full LSSTCam field of view.}
\label{tab:observability_date}
\end{table}

\section{Small-area target fields}

We separately present our target requests for small area ($\lesssim10$~deg$^2$) and wide area fields.  Small area targets may be scheduled throughout the commissioning period, while wide area observations would most likely be scheduled by the commissioning team in the context of the science verification surveys.  Here we present our strategy related to small area fields.  Our preferences for wide area observations are addressed in Section~\ref{sec:WFD} below.

\subsection{Target classes and scientific categories}
\label{sec:classes_categories}

The commissioning schedule remains uncertain, as does the delivered data quality that will be achieved at various stages of system integration.
Our prioritization of the observing fields and/or science goals may change based on the quality of the data, the final commissioning schedule, and amount of time on sky.  Accordingly, in the summaries below, we have categorized the candidate target fields into three classes to indicate both the quality and volume of data that would be optimal to perform science validation studies from the DESC perspective. These three target classes approximately map to three types of observing campaigns that are being considered by the Rubin Observatory Project.

\begin{itemize}
    \item \textbf{Class 1 targets} are those that could be observed opportunistically during the assembly, integration, and testing phases with ComCam and LSSTCam. During these engineering phases, windows during which the system configuration approaches nominal science imaging quality may be anticipated. For these targets, even limited amounts of data (e.g., an hour) with below-average image quality (e.g., 1.4 arcsec) and a limited number of bands and unpredictable cadence, would still be useful to test DESC analysis pipelines. $\sim80$ visits for a given field could be acquired in roughly 1 hour.
    \item \textbf{Class 2 targets} are well matched to observing campaigns that would achieve the equivalent exposure of 1 year of the Wide-Fast-Deep survey in multiple bands, with nominal data quality. 
    For example, it would be possible to acquire 100 total visits, or $\sim20$ visits in each of 5 filters for LSSTCam, in $\sim80$ minutes wall time. 
    Note that ComCam only has 3 filters loaded at a time.
    \item \textbf{Class 3 targets} are best suited for deep drilling observations with nominal data quality that would reach integrated exposures equivalent to or exceeding the planned 10-year depth of the WFD survey in as many bands as possible ($\gtrsim800$ visits, or roughly 9 hours).
\end{itemize}

We also identify example scientific validation studies that can be performed with each candidate target field in the broad science categories below.

\begin{itemize}
    \item \textbf{Astrometric Calibration (A):} includes dense stellar fields for star flats;
    \item \textbf{Astrophysical Target of Interest (I):} includes rich galaxy clusters, dense stellar fields suitable for microlensing;
    \item \textbf{Photometric Calibration (P):} includes spectrophometric standards;
    \item \textbf{Photometric Redshift (Z):} includes fields with deep spectroscopy and/or multiband imaging extending into the infrared and ultraviolet;
    \item \textbf{Time Domain (T):} includes the LSST DDFs and other high-cadence fields;
    \item \textbf{Space-based Imaging (S):} includes fields observed with HST.
\end{itemize}

These observation classes and scientific categories are listed for our highest priority fields in Table~\ref{tab:priority}, and are available digitally for all fields at the resources listed in Appendix~\ref{sec:Appendix-Accompanying-Material}.

\subsection{Field Prioritization}

Our single highest priority is being able to build good templates in the DDFs during commissioning so that SN science operations can begin in LOY1, operating in a mode similar to DES/SNLS/PS1 MDS. 
There are four approved DDFs, and Euclid Deep Field South (EDFS) has been proposed as a fifth DDF (see Appendix~\ref{sec:Appendix-field-list}).
The DDFs overlap multiple external datasets that are useful for a broad range of science validation studies. Deep imaging in these fields could be used to both investigate subtle systematic effects and calibrate the performance of the wide-field observations.

Given that observing time during commissioning will be limited, in addition to presenting all of our requested fields, we have also curated a set of highest-priority targets ordered by when they are visible from Cerro Pachon.  The information is taken from Table~\ref{tab:observability_date}. In each observing period, priority fields were selected by maximizing the number of DESC analysis working groups that indicated their importance. We did not explicitly attempt to balance the tests we could do using the set of high-priority fields. In Table~\ref{tab:priority}, fields are listed roughly in order of descending priority.  The table is also subdivided by observing quarter. Within each quarter, we list high priority and backup fields.  The fields are  indicated with their observing class and scientific category in order to facilitate the commissioning team finding appropriate fields based on need and opportunity.  Again, these are only a subset of the fields we propose in Table~\ref{tab:observability_date} and they do not include the wide-area fields discussed in Section~\ref{sec:WFD}.  As some fields can be observed with ComCam, while others require the entire field of view of LSSTCam, a required field of view for each field is also indicated in Table~\ref{tab:observability_date}.

\begin{table}[!htbp]
\centering
\begin{tabular}{llll}
 \toprule
Priority & Field & Category & Class \\
\midrule
\multicolumn{4}{c}{\textbf{January - March}} \\
\midrule
\textit{high} & LSST DDF COSMOS &  \textbf{P S T Z} & 3, 1 \\
& MS0451-03 &  \textbf{I S Z$^a$} & 2 \\         
& A901/902 &  \textbf{I S} & 2 \\  
\midrule
\textit{back-up} & LSST DDF CDFS &  \textbf{I S T Z} & 3, 1 \\ 
& SXDS / UKIDSS UDS / C3R2 &  \textbf{I S Z} & 3 \\ 
\midrule
\multicolumn{4}{c}{\textbf{April - June}} \\
\midrule
\textit{high} & LSST DDF COSMOS &  \textbf{P S T Z} & 3, 1 \\          
& A901 / 902 &  \textbf{I S} & 2 \\  
\midrule
\textit{back-up} & LSST DDF ELAIS-S1 &  \textbf{T Z} & 3, 1 \\ 
& A1689 &  \textbf{I S} & 2 \\ 
& DES SN S1 / S2 &  \textbf{Z} & 3 \\
\midrule
\multicolumn{4}{c}{\textbf{July - September}} \\
\midrule
\textit{high} & LSST DDF XMM-LSS / VVDS-Deep / VIPERS  &  \textbf{Z T} & 3, 1 \\ 
& LSST DDF ELAIS-S1 &  \textbf{T Z} & 3, 1 \\ 
&  SXDS / UKIDSS UDS / C3R2 &  \textbf{I S Z} & 3  \\
\midrule
\textit{back-up} & A1835 &  \textbf{I S  } & 2 \\ 
& A2390 &  \textbf{I S} & 2  \\ 
& DEEP2-2h Field 4 &  \textbf{P S T Z} & 3, 1 \\  
\midrule
\multicolumn{4}{c}{\textbf{October - December}} \\
\midrule
\textit{high} & LSST DDF CDFS &  \textbf{I S T Z} & 3, 1 \\
& MS0451-03 &  \textbf{I S Z$^a$} & 2 \\   
&  LSST DDF XMM-LSS / VVDS-Deep  &  \textbf{T Z} & 3, 1 \\ 
\midrule
\textit{back-up} & SXDS / UKIDSS UDS / C3R2 &  \textbf{I S Z} & 3 \\
& Euclid Deep Field South &  \textbf{S T Z} & 1 \\ 
& RX J2248.7-4431 &  \textbf{I S Z$^a$} & 2 \\ 
 \bottomrule
 \end{tabular}
 \caption{Prioritized fields, per quarter. See Section~\ref{sec:classes_categories} for definition of classes and scientific categories. $^a$ Cluster fields are marked with ``Z'' if $\gtrsim 2500$ spec-z's are available, extending to faint targets. These should be considered photo-z testing fields, not training fields.}
 \label{tab:priority}
\end{table}

\begin{figure}[!htbp]
  \begin{center}
    \vspace{0.25in}
    \includegraphics[width=6.0in]{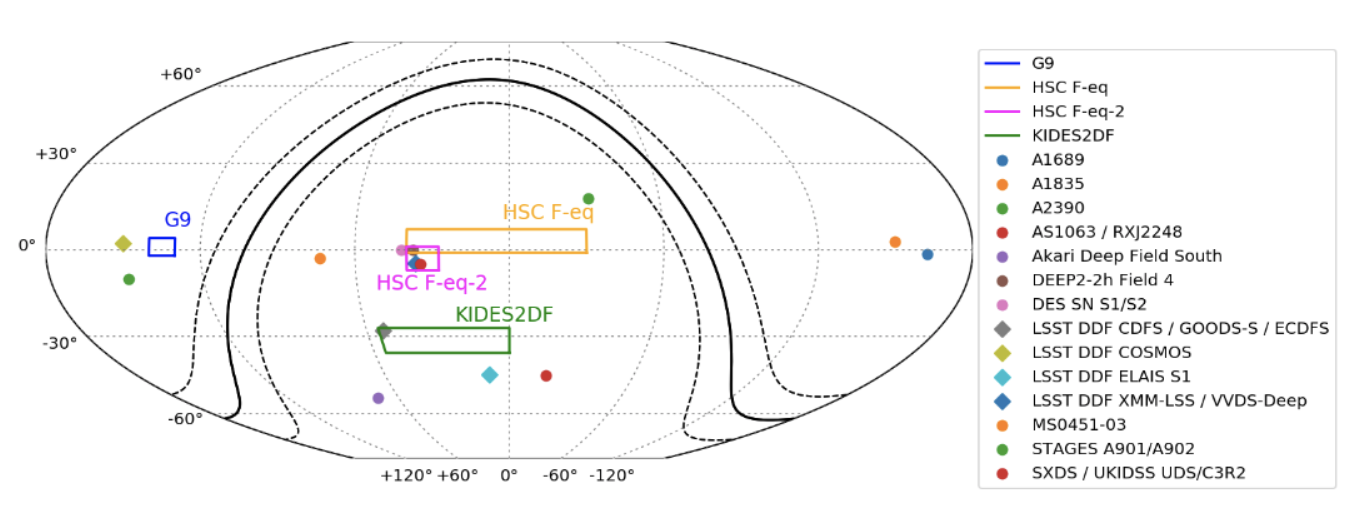}
    \vspace{0.02in}
    \caption{Locations of the priority fields summarized in Table~\ref{tab:priority} are shown, alongside the locations of the four wide-area target regions.}
    \label{fig:field-priority}
  \end{center}
\end{figure}

Figure~\ref{fig:field-priority} is a visual representation of the field centers of these high-priority suggestions.   The locations of several wide-area regions of interest coinciding with other survey footprints are also shown.  This information is further presented graphically in Figure~\ref{fig:visibility}.

\subsection{Dithering tests for DDF fields}
\label{sec:DDF-tests}

Rubin commissioning offers an opportunity to test shifts in telescope pointings (dithers) that will be utilized during the LSST. We assume that the DDFs will utilize random rotational dithers in order to move objects to as many different positions within the focal plane as possible without enlarging the deep field, which would result in a loss of depth.  However, some types of imaging artifacts cannot be eliminated through purely rotational dithers, leading to a need for additional translational dithers of an undetermined amplitude even in DDFs.  

For translational dithers in a commissioning DDF, we recommend comparing the efficacy of sensor-sized random dithers vs. raft-sized random dithers vs. no dithers, with random rotational dithers still applied in all three cases.  The motivation is to see what size translational dithers are needed in DDFs to remove both optical ghosts and scattered light off of surfaces other than the optical system~\citep[for further discussion of the removal of such artifacts, see][]{ 2016A&C....16...67D,2014PASP..126..158G, 2016MNRAS.460.2245J}.  
The rationale for including   DDF pointings without  translational dithers as one of the three options described above is that the overlapping WFD pointings will themselves have large translational dithers, and it is possible that the combined imaging will be able to eliminate the artifacts.  Hence we request that a realistic ratio of WFD-like and DDF-like observations be taken at a DDF location during commissioning, with the DDF-like observations utilizing an equal mix of the three translational dither patterns mentioned above.  The metric for success is the level of reduction of those artifacts in stacked images from each dithering scheme, after combination with the WFD-like images and DM processing.

\section{Considerations for Wide-Area Fields}
\label{sec:WFD}

From the DESC perspective, a smaller area observed with at least three bands is preferable to a single filter over a large area. Single-band observations make it impossible to measure photometric redshifts that are needed for most dark energy probes. Three bands would allow for color-color calibration based on stellar populations (i.e., stellar locus regression).  Four or more bands would be more suitable for evaluating photometric redshifts in the wide field and for evaluating the recovered density of specific galaxy populations used in large-scale structure analyses.
Furthermore, templates in a single filter are not useful for SN operations as the lack of color information makes identification and classification impractical.  Tests of clustering-based calibration of photometric redshifts depend on overlap with existing wide area spectroscopic surveys such as DESI or SDSS Stripe 82, so areas of the sky with existing redshift coverage should be prioritized, in case at least 4 filters can be used to build these estimates.
Two filters chosen in a moon phase-dependent pattern ($g$+$r$ in dark to 50\% grey, r+i otherwise) is the minimum that is useful for the SN WG commissioning goals. This strategy would allow testing of operations in two sets of pairs. 

The DESC would prefer wide-field observations in a long contiguous strip overlapping with existing datasets, for example, the equatorial Stripe 82 region.
An alternate choice would be imaging large contiguous regions surrounding the DES S1, S2, S3, or the HSC SN fields.
Having a subset of Stripe 82 ($\sim$100--200~deg$^2$) imaged in [$u$]$gri$[$z$] during commissioning for validation of the WFD operations prior to LOY1 would be extremely useful for SN and SL, and would be of use to the PC WG and other groups as well. 
Regularly cadenced observations in the same filter(s) spaced by 1 to 4 nights are ideal for science validation using SN lightcurves.

It should be noted that we do not want observations aimed at building templates for a WFD region to detract from other validation activities in commissioning (particularly testing prompt processing, alert distribution, and difference image analysis with a sufficient number of visits in a smaller sky area).

The current scheduler implements a baseline plan for translational dithers in the WFD survey based upon re-tessellating a hexagonal tiling of the sky each night.  This equates to a random-per-night translational dithering scheme, as recommended by \citet{2016ApJ...829...50A}.   
For added rotational dithers in a commissioning WFD region, we propose to test the effectiveness of returning to a single randomly chosen nightly value of rotSkyPos (modulo 180 degrees) during each slew and after each filter change or cable unwrapping.  That can be tested vs. an approach that utilizes a single randomly chosen value of rotTelPos each night at the start of each visit, except during re-visits so that visit pairs are performed at the same rotSkyPos.  rotTelPos must change slowly during exposures to maintain rotSkyPos, and the hope is that achieving a nearly uniform distribution of rotSkyPos over many nights will also generate a nearly uniform distribution of rotTelPos.  
The results should be evaluated by looking for evidence of PSF residuals in stacked, DM-processed images that mimic shear and quantifying how much these are reduced using each strategy. An additional metric is the Kuiper test, which can be used to evaluate the uniformity of the distribution of rotational angles \citep[see][]{2020arXiv200612538A}.  
Figure~\ref{fig:field-priority} shows the footprints of several relevant wide-area fields.

\section{Spectrophotometric Standards}

As described above, we propose to interleave commissioning observations of the DDFs
with observations of selected spectrophotometric standards. The goal is to secure an early flux calibration of sky patches observed during commissioning, in particular the DDFs, which are of primary importance for supernovae.  Beating down the observation noise affecting the calibration metrology chain requires accumulating many visits to the science fields and calibration targets. Commissioning observations that frequently revisit the same fields offer a good opportunity to achieve this.  From our experience with wide-field surveys, we assume a night-to-night repeatability of 1\% (10 mmags). In order to average the results down to 0.1\% (1 mmag) calibration performance, we estimate that on the order of 100 visits separated by at least 30 minutes each are necessary.

This challenging goal can be attained, provided that several visits to spectrophotometric standards are scheduled during each night, interlaced with the DDF/Wide commissioning targets.

\begin{figure}[!htb]
  \begin{center}
   \includegraphics[width=5.0in]{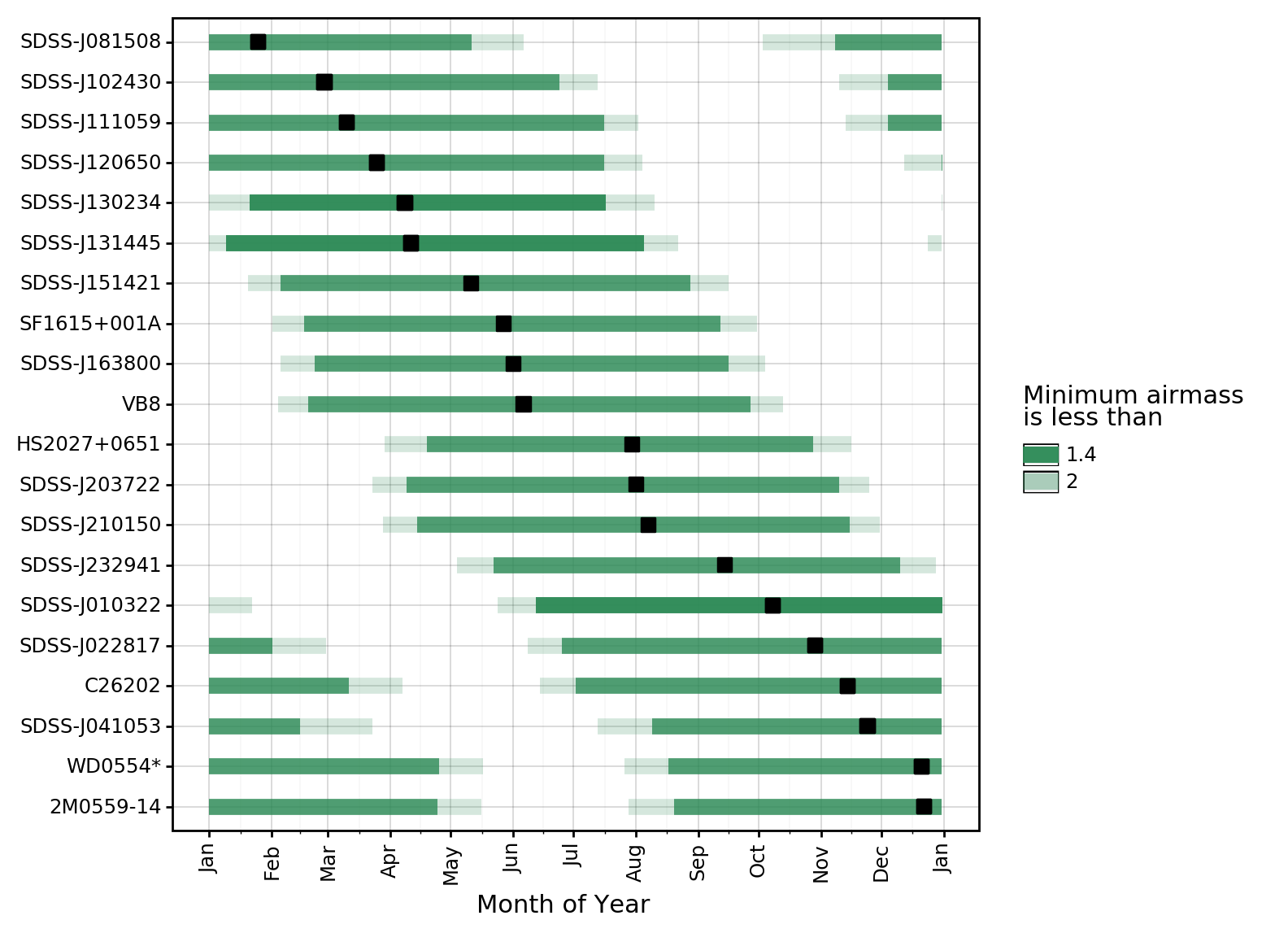}
    \caption{Visibility as a function of month for the spectrophotometric targets in Table~\ref{tab:spectroscopic_standard_star_observability} sorted by optimal yearly observing date. For each target, green and light green bands show when the target has air-mass less than 1.4 (solid) or 2.0 (light) at the time of optimal observing. The optimal observing date for each target is indicated with a black square.}
    \label{fig:spectrophotometric-visibility}
  \end{center}
\end{figure}

\begin{figure}[!htb]
  \begin{center}
     \includegraphics[width=5.0in]{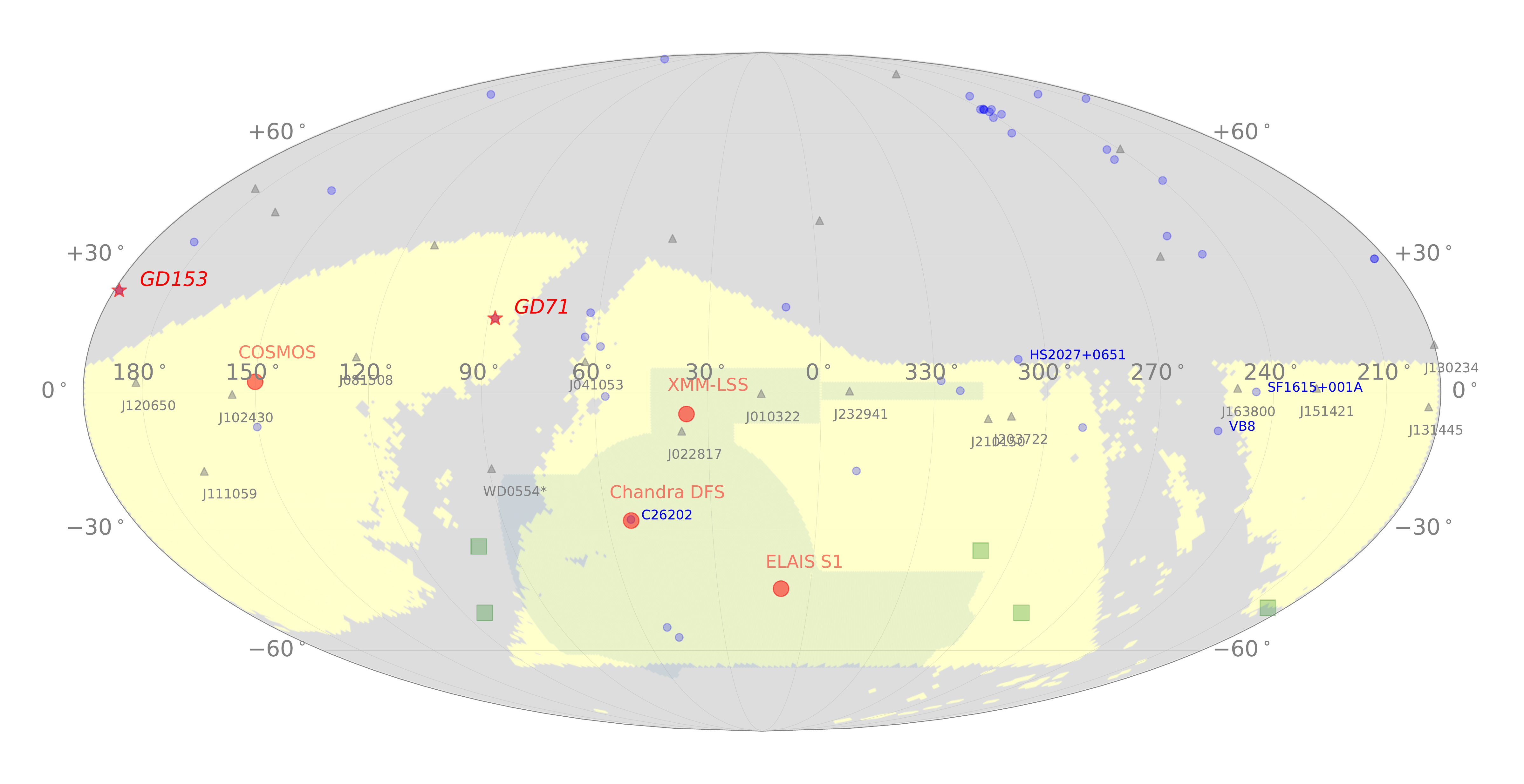}
     \caption{Calibration targets: LSST DDF pointings (red dots),
       CALSPEC stars (blue dots), \citet{2019ApJS..241...20N} standard stars (gray upper triangles). The primary CALSPEC white dwarf references within reach of LSST are represented with red stars.  The green squares
       are the DES starflat fields and the shaded areas correspond to the footprint of LSST (yellow) and DES (primarily light green where overlapping the yellow region).}
    \label{fig:calibration_targets}
  \end{center}
\end{figure}

Table~\ref{tab:spectroscopic_standard_star_observability} lists our current selection of spectrophotometric standards. Their visibility as a function of the time of the year is shown in Figure~\ref{fig:spectrophotometric-visibility}. Their distribution in equatorial coordinates is shown in Figure~\ref{fig:calibration_targets} as well.   All are anchored on the white dwarf flux scale \citep[e.g.,][and references therein]{2014PASP..126..711B} which is currently the state of the art regarding relative flux calibration. 
The objects listed in Table~\ref{tab:spectroscopic_standard_star_observability} are a subset of either the CALSPEC library or one of its extensions developed by \citet{2019ApJS..241...20N}.  
We selected objects which (1) have declination below $+10$ deg, i.e., are observable from Cerro Pachon at a moderate airmass, and (2) are fainter than $V \sim 15$, which means that they will not saturate on standard 15~s LSST exposures.

\begin{table}[!htbp]
\tiny
\begin{tabular}{lrrccccccr}
\toprule
        Star name & RA (J2000)    & DEC (J2000)    &   Optimal Date & Optimal Airmass & $T_{1.4}$ & $T_{2.0}$ & Dates $\sec(z) < 1.4$ & Dates $\sec(z) < 2.0$ & Magnitude \\
                  & (deg) & (deg)  &                &                 & (hr) & (hr) & & & (mag) \\
\midrule  
                                       SF1615+001A & 244.56 & 0.00 & May-27 (147) & 1.16 & 4.5 & 7.2 & Feb-17 to Sep-12 & Feb-01 to Sep-30 & 16.75 ($V$) \\
                                               VB8 & 253.90 & -8.39 & Jun-06 (157) & 1.08 & 5.4 & 7.8 & Feb-19 to Sep-27 & Feb-04 to Oct-13 & 16.92 ($V$) \\
                                       HS2027+0651 & 307.39 & 7.02 & Jul-30 (211) & 1.26 & 3.3 & 6.3 & Apr-19 to Oct-28 & Mar-29 to Nov-16 & 16.9 ($V$) \\
                                            C26202 & 53.14 & -27.86 & Nov-14 (318) & 1.00 & 6.7 & 7.4 & Jul-02 to Mar-11 & Jun-14 to Apr-07 & 16.64 ($V$) \\
                                         2M0559-14 & 89.83 & -14.08 & Dec-22 (356) & 1.04 & 5.9 & 6.7 & Aug-20 to Apr-24 & Jul-28 to May-16 & 13.80 ($J$) \\
\midrule
                                      SDSS-J081508 & 123.79 & 7.53 & Jan-25 (025) & 1.26 & 3.2 & 6.4 & Nov-08 to May-11 & Oct-03 to Jun-06 & 20.33 (PS1 $r$) \\
                                      SDSS-J102430 & 156.13 & -0.54 & Feb-27 (058) & 1.15 & 4.5 & 7.3 & Dec-04 to Jun-24 & Nov-10 to Jul-13 & 19.29 (PS1 $r$) \\
                                      SDSS-J111059 & 167.75 & -17.17 & Mar-10 (069) & 1.03 & 6.2 & 8.6 & Dec-04 to Jul-16 & Nov-13 to Aug-02 & 18.30 (PS1 $r$) \\
                                      SDSS-J120650 & 181.71 & 2.03 & Mar-25 (084) & 1.18 & 4.2 & 7.0 & Dec-31 to Jul-16 & Dec-12 to Aug-04 & 19.10 (PS1 $r$) \\
                                      SDSS-J130234 & 195.64 & 10.21 & Apr-08 (098) & 1.31 & 2.5 & 6.1 & Jan-21 to Jul-17 & Dec-31 to Aug-10 & 17.49 (PS1 $r$) \\
                                      SDSS-J131445 & 198.69 & -3.24 & Apr-11 (101) & 1.12 & 4.9 & 7.5 & Jan-09 to Aug-05 & Dec-24 to Aug-22 & 19.55 (PS1 $r$) \\
                                      SDSS-J151421 & 228.59 & 0.80 & May-11 (131) & 1.17 & 4.4 & 7.0 & Feb-05 to Aug-28 & Jan-20 to Sep-16 & 16.10 (PS1 $r$) \\
                                      SDSS-J163800 & 249.50 & 0.79 & Jun-01 (152) & 1.17 & 4.4 & 7.0 & Feb-22 to Sep-16 & Feb-05 to Oct-04 & 19.31 (PS1 $r$) \\
                                      SDSS-J203722 & 309.34 & -5.22 & Aug-01 (213) & 1.10 & 5.0 & 7.6 & Apr-09 to Nov-10 & Mar-23 to Nov-25 & \\
                                      SDSS-J210150 & 315.46 & -5.76 & Aug-07 (219) & 1.10 & 5.2 & 7.6 & Apr-14 to Nov-15 & Mar-28 to Nov-30 & 19.05 (PS1 $r$) \\
                                      SDSS-J232941 & 352.42 & 0.19 & Sep-14 (257) & 1.16 & 4.4 & 7.1 & May-22 to Dec-10 & May-04 to Dec-28 & 18.45 (PS1 $r$) \\
                                      SDSS-J010322 & 15.84 & -0.35 & Oct-08 (281) & 1.15 & 4.5 & 7.2 & Jun-12 to Dec-31 & May-24 to Jan-22 & 19.57 (PS1 $r$) \\
                                      SDSS-J022817 & 37.07 & -8.45 & Oct-29 (302) & 1.08 & 5.5 & 7.6 & Jun-25 to Feb-01 & Jun-08 to Feb-28 & 20.19 (PS1 $r$) \\
                                      SDSS-J041053 & 62.72 & 6.51 & Nov-24 (328) & 1.25 & 3.4 & 6.5 & Aug-09 to Feb-15 & Jul-13 to Mar-23 &  \\
                                           WD0554* & 89.26 & -16.59 & Dec-21 (355) & 1.03 & 6.1 & 6.7 & Aug-17 to Apr-25 & Jul-26 to May-17 & \\
\bottomrule
\end{tabular}
\caption{Table for spectrophotometric standard stars observability
  throughout the calendar year. The upper part of the table contains the CALSPEC stars fainter than $V\sim 15$ and at a declination $\delta < +10$. The lower part  of the table contains the fainter spectrophotometric standards from \citet{2019ApJS..241...20N} -- also anchored on the HST WD flux scale.  $T_{1.4}$ ($T_{2.0}$) represent the time in hours that a
  given field is observable at an airmass less than 1.4 (2.0) on the optimal night of the year.  The two rightmost columns provide the   range of calendar dates for which the field is observable at some   point during the night (solar elevation angle $< -18^{\circ}$) at an  airmass $\sec(z)$ below the indicated threshold.} 
  \label{tab:spectroscopic_standard_star_observability}
\end{table}

The selection presented in Table~\ref{tab:spectroscopic_standard_star_observability} will be refined as the commissioning program is finalized, and as fainter, high-quality spectrophotometric standards are released. 
We note that the number of CALSPEC stars that can be observed in this program is limited: most are in the northern hemisphere, and brighter than $V\sim 14$. 
If necessary (e.g., if tensions are detected between calibrations from different spectrophotometric objects), it is possible to increase the number of bright CALSPEC objects ($V > 12$) in the calibration program by decreasing the exposure time of the calibration observations down to 1~s (in which case we recommend taking 1~s and 15~s exposures back to back for intercalibration of the short- and long-duration exposures). Finally, we note that AuxTel observations will allow an early test of this program, notably by comparing spectrophotometric standard stars and detecting potential tensions between them.

\section{Acknowledgments}

The authors would like to extend special thanks to our internal DESC reviewers Andrew Connolly of the University of Washington and David Kirkby of the University of California at Irvine for their careful reading of the manuscript along with Anupreeta More of the Indian Inter-University Centre for Astronomy and Astrophysics who provided suggestions of double source-plane lens targets.

The DESC acknowledges ongoing support from the Institut National de Physique Nucl\'eaire et de Physique des Particules in France; the Science \& Technology Facilities Council in the United Kingdom; and the Department of Energy, the National Science Foundation, and the LSST Corporation in the United States.  DESC uses resources of the IN2P3 Computing Center (CC-IN2P3--Lyon/Villeurbanne - France) funded by the Centre National de la Recherche Scientifique; the National Energy Research Scientific Computing Center, a DOE Office of Science User Facility supported by the Office of Science of the U.S.\ Department of Energy under Contract No.\ DE-AC02-05CH11231; STFC DiRAC HPC Facilities, funded by UK BIS National E-infrastructure capital grants; and the UK particle physics grid, supported by the GridPP Collaboration.  This work was performed in part under DOE Contract DE-AC02-76SF00515.

\clearpage
\bibliography{references}

\appendixpage
\appendix

The following appendices contain more detailed explanations of the physics drivers behind the selection of fields, a detailed list of all selected fields including relevant references, and further information on how the reader can access the spreadsheets and tools used to produce this document.

\section{Science drivers for the targets from the DESC}
\label{sec:Appendix-Science-Drivers}

Each science working group in the DESC has particular needs related to testing their pipelines and ensuring they are ready to achieve DESC science objectives.   As such each working group has specific and sometimes unique needs for test/validation data from early LSST observations.  Here, we outline the scientific themes of the  working groups along with the classes of tests they will employ and what that implies for needed fields.

\textbf{Calibration:} DESC has stringent requirements on photometric calibration, most of them driven by SN~Ia cosmology. The critical requirements are on the relative (band-to-band) flux calibration which needs to be controlled at the 0.1\% level, and on the metrology of the survey passbands, with a targeted accuracy of 1\AA~\citep{2018arXiv180901669T}. The sources of calibration uncertainty can be categorized in four main groups: (1) external uncertainties (i.e., the uncertainty on the primary flux references), (2) internal uncertainties from the photometry chain that connects the flux standards and the science observations, (3) uncertainties on the determination of the effective survey passbands, and (4) determination of the atmospheric transmission with AuxTel. The external uncertainties are addressed by dedicated projects such as starDICE, SCALA or NISTstars. The main goal of commissioning is to evaluate the contribution of the photometric chain to the total error budget (group 2).   The Photometric Calibration WG proposes to use the repeated DDF observations to assess the photometric repeatability and the sensitivity of the point-source flux estimators to the observing conditions. Another goal is to evaluate the detrending process and the uniformity of calibrated fluxes from dithered observations of moderately dense fields, as well as the uniformity reached on a relatively larger ($\sim$ 100--200~deg$^2$) WFD region with ubercal/FGCM-like methodologies. Finally, we propose to undertake an early transfer of the CALSPEC flux scale to the DDFs by inserting observations of spectrophotometric standards in the observation program shortly before and after the DDF observations.

\textbf{Clusters:} The main cosmological probe with clusters of galaxies for LSST will be measurements of the shape and evolution of the halo mass function.  LSST will find clusters via their galaxy overdensities, and use weak lensing to estimate their masses. For commissioning, the Clusters (CL) WG is interested in testing systematics related to WL shape measurements and photometric redshifts, particularly in overdense fields of galaxies, where blending is an increased concern. The fields suggested by CL have existing HST imaging to test the performance of the LSST deblender and shape measurements in comparison to space-based imaging.  In addition, these fields have comprehensive spectroscopic coverage (and/or imaging in a large number of bands for accurate and precise photo-$z$s) which will enable assessing the quality of LSST photo-$z$ estimates. Commissioning imaging in the full filter set ($ugrizy$), ideally to 10-yr survey depth, would enable these tests.  The fields listed here all have Chandra and/or XMM X-ray imaging, and most have SZ data, anticipating synergy with eROSITA as well as Simons Observatory and CMB-S4.  Many clusters also have NIR and/or Spitzer imaging, which can be used to further improve photometric redshifts.

\textbf{Dark Matter:} The Dark Matter (DKM) WG seeks to understand the fundamental nature of dark matter through observations of both the local and distant universe. Many of the targets of DKM studies (galaxy clusters, dwarf galaxy weak lensing, ultra-faint Milky Way satellites, etc.) have very similar survey requirements to the other WFD components.  The interests of DKM overlap with those of the WL working group in the need for good image quality (e.g., for star/galaxy separation and strong lens identification) and shares requirements for uniformity with LSS. The main area in which DKM needs diverge from other DESC WGs is the desire to do difference imaging in dense stellar fields to search for microlensing events. Thus, the distinct DKM interest for commissioning is the observation of dense stellar fields with the highest feasible cadence (sub-minute repetition rate) and 2--3 day cadence (as well as longer time baselines if possible). The DKM WG proposes three fields overlapping the LMC, SMC, and Galactic Bulge, which overlap with existing MACHO, EROS, and DECam observations. These cover a range of different RAs, but one target would be enough.

\textbf{Large-Scale Structure:} This probe is sensitive to any effect (e.g., astrophysical or observational) that will introduce spurious correlations in the inferred spatial distribution of a given galaxy population at a given redshift.
Usage of well-characterized populations from previous surveys, such as Luminous Red Galaxies, will be important for understanding the accuracy of each component contributing to the angular power spectrum.  The statistical errors in the DES Science Verification data set with $\sim$200~deg$^2$ allowed the measurement of clustering up to interesting cosmological scales of a few degrees. Therefore, the Large-Scale Structure (LSS) WG is interested in tests performed on a field (or a small number of them) that is hundreds of square degrees in scale, overlaps previous surveys, and is observed to the equivalent of 1-year depth for the LSST. In addition, 10-year depth observations of spectroscopic fields and spectrophotometric calibrators will be needed for photo-$z$s and understanding photometric biases, as detailed below, and for understanding the effect of deblending on the determination of large-scale structure with 10-year depth observations.

\textbf{Photometric Redshifts:} DESC has stringent requirements on photo-$z$ performance~\citep{2018arXiv180901669T}, and meeting those requirements will depend critically on the availability of objects with secure redshifts for both training of the algorithms and calibration of residual systematics.  Given that so much of DESC science depends on photo-$z$ quality, verifying that the photo-$z$ requirements can be achieved is a high priority for the Photo-Z WG during commissioning.  The specific issues of training and calibration are discussed in detail in \citet{Newman:15}, but in summary a training set of galaxies is required that is as close as possible to spanning the observed galaxies' range of apparent magnitude and colors in order to define the mapping from fluxes to redshift, including degenerate areas of parameter space.  Targeted observations to verify photometric repeatability and photometric spatial uniformity are essential, and already covered as part of planned commissioning observations~\citep{LSE-79}.  As photo-$z$ estimates are highly dependent on photometric errors, full 10-year depth observations in all six filters covering key fields with existing deep spectroscopic samples will be needed.  We do not expect to obtain fully spanning spectroscopic samples, and thus need to assume that some systematic biases in redshift will remain.  Ideally, these biases could be calibrated using angular clustering-based techniques \citep[e.g.,][]{Newman:08}.  Those methods require secure spectroscopic redshifts over large areas ($\gtrsim$ hundreds to thousands of square degrees) from surveys such as DESI to overcome uncertainties due to sample variance. For this reason, observations that overlap with large spectroscopic surveys (O(1000)~deg$^2$) that enable clustering-based calibration tests are desirable during commissioning observations in order to test the framework for such calibrations during the main survey. In addition to the major issues of training and calibration, complications such as star/galaxy separation and source blending will contribute to photo-$z$ systematics. The availability of high resolution space-based imaging in at least one of the spectroscopic training fields observed will be an additional benefit to photometric redshift studies during commissioning.

\textbf{Strong Lensing:} The strong lensing (SL) WG seeks to study the cosmic expansion history through (1) the measurement of time delays of multiply-imaged variable and transient objects, and (2) the measurement of ratios of Einstein radii in compound lens systems. The ability to both find and characterize strong lens systems in LSST will be particularly sensitive to the distribution of delivered image quality (including the bluer bands) since the light of the foreground lens and background source(s) is blended for most galaxy-scale lenses. Searches for strong lenses in LSST might be improved by using subsets of images that have been selected and weighted to optimize the effective coadded image quality. 
For many of the compound lenses, however, utilizing the deepest possible coadded images would increase the efficiency of finding faint lensed features arising from more distant sources. Commissioning observations of compound lenses (preferably at 10-year depth) would be particularly useful to test the efficiency of various lens detection algorithms.
Studies of strongly lensed quasars and transients additionally require accurate difference image photometry to measure time delays.
Commissioning data for several lensed quasar systems with a range of angular separations, including those in the range where blending due to typical seeing conditions is expected, would be valuable for testing the detection of such systems and building reliable light curves of the multiple quasar images (though not necessarily extracting the time delays from the light curves).
The signature of strongly lensed transients in the alert stream is of particular interest given needs for rapid identification and follow-up.
Lensed SNe candidates, if any, would more likely come from a wide-area part of the science verification surveys (at least hundreds of square degrees) with at least some cadence, and less likely from the high-cadence and small area part of the science verification surveys that go to 10-year or 20-year depth.
A goal is to understand how the difference image pipeline performs for transients close to the cores (within an arcsecond) of massive elliptical galaxies.

\textbf{Supernovae:} This probe relies on comparing the brightnesses of distant supernovae with those in the nearby Universe, and is therefore sensitive to any effect that impacts the determination of brightness, particularly effects that evolve with redshift. LSST will find more supernovae at higher redshifts than any previous survey. For commissioning, the Supernova (SN) working group is interested in imaging of the deep-drilling fields (DDFs) in order to (1) build source catalogs prior to the survey start and determine the redshifts of potential host galaxies, (2) verify and validate the pipeline processing and alert production, and (3) generate templates for difference image analysis (DIA) when the main survey starts. We anticipate that supernovae discovered in the DDFs  will be critical to early cosmological studies by the SN WG, as the light curves will be better sampled, and the fields have significant ancillary data, at least over part of each field. 

Therefore, very deep imaging $gri$ (10 $\times$ nightly depth on a single photometric night with good seeing) of the DDFs are of highest priority to generate templates for difference imaging and source catalogs for these fields for use during LOY1. Imaging of non-DDFs that have been used for SNe science (e.g., PS1, DES and SNLS) is useful for cross-calibration, but not as critical as pre-imaging of the DDFs as the latter also encompass some DES and SNLS fields. Deep $gri$ (5 $\times$ nightly depth on a single photometric night with good seeing) imaging to generate templates and source catalogs of a reasonably large ($\approx$100--200~deg$^2$) region is very desirable for verification and validation of LSST wide-fast-deep (WFD) survey operations for the SN WG. 

Finally, we also seek some regularly spaced (with at least a 4 day cadence in the same filters, though faster is better) time-domain observations of the DDFs and the 100--200~deg$^2$ region.  These would be valuable to detect time-variable sources during commissioning itself and test the SN pipeline from end to end. These images can be taken in any seeing conditions with airmass $<$ 2, with standard exposure times. We do not specify a total number of observations needed with the above frequency as the total duration of commissioning is unknown at this juncture, but will accept as many as is feasible.

We anticipate that commissioning imaging desired by the LSS WG will also be suitable for SN WG needs, provided that repeated observations in the same filter(s) are taken with at least 1-day between them; ideally the cadence would match that of the WFD survey. Stripe 82 would be an excellent choice for this contiguous region. 

\textbf{Weak Lensing:} This cosmological probe requires detecting any effect (astrophysical or systematic) that creates coherent distortions in the shapes of galaxy populations at similar redshifts. Accurate measurements require the calibration of both the redshifts and the shears.   The former relies on deep observations of fields with existing high-quality observations from other surveys  that are spectroscopic or in complementary photometric bands. The latter is facilitated through statistical comparisons with previously calibrated shear fields, from previous weak-lensing (WL) surveys. As such, the WL working group (WG) would like commissioning-era observations that enable comparative tests using large contiguous areas that overlap multiple previous WL surveys with at least 1-year depth in LSST, which will be comparable to the usable depth of the calibrated pre-cursor surveys.  Experience with the DES Science Verification Data demonstrated that field sizes of a few-hundred square degrees were sufficient to explore many subtle systematic effects in the data.  Overlap with smaller footprint space-based data-sets will likely be useful all the way to LSST 10-year depth as their increased signal-to-noise ratio can compensate the relatively poorer angular resolution of the LSST survey.  Additionally, some fields with spectroscopic data should be observed at full 10-year depth, with a typical distribution of filters, to characterize photometric redshifts.

\section{Collated detailed descriptions of fields}
\label{sec:Appendix-field-list}

In this appendix, each field referred to in the tables and figures is listed in alphabetical order.  The rational and use for each field is given along with relevant pre-cursor datasets and references. Scientific category codes are listed beside the field name (see Section \ref{sec:classes_categories}). The names of the DESC working groups who requested the fields are also indicated with the coloured flags. These groups include Clusters (CL), Dark Matter (DKM), Large Scale Structure (LSS), Photometric Calibration (PC), Strong Lensing (SL), Supernova (SN) and Weak Lensing (WL). The number of working groups requesting a field was used in the prioritization process. Note that some entries in this table (e.g., clusters of galaxies) contain multiple related fields.


\begin{itemize}

\item  \textbf{A901/902} (IS) \\ \votecl\votewl \\
Overlapping data: HST, 2dF, Spitzer, XMM-Newton, GALEX, GMRT  \\  
The A901/902 super-cluster ($z=0.165$) was imaged with an 80-tile HST/ACS mosaic in the F606W band as part of the STAGES survey \citep{gray09}.  It was part of the COMBO-17 survey, and was imaged in 17 broad and medium bands with WFI on the 2.2m telescope, providing high-quality photometric redshifts for 15000 galaxies to $R<24$. 2df spectroscopy, GALEX, Spitzer, XMM and GMRT data are also available.

\item \textbf{AdvACT clusters} (IS) \\ \votecl \\ 
We add a number of clusters from the AdvACT SZ survey.  Most of these are very well known and extensively studied, with generally a wealth of multi-wavelength data.  All clusters here have been imaged with HST, and were targeted with MUSE IFU spectroscopy (providing deep spectroscopy for {\it every} object in a $1^{\arcmin} \times 1^{\arcmin}$ field.
These include A68 (24 min MUSE; 1 pointing in F110W, F160W, F132N, F814W), A1689 \citep[2.5 hr MUSE, extensive HST observations][]{MUSE_A1689}, A1835 (23 min MUSE, 1 HST pointing in 7 ACS+WFC3 bands plus NIC2 data), and A2390 (30 min MUSE, 1 HST pointing in F125W, F126W, F850LP; plus WFPC2 data), all within the AdvACT field; RCS2 0327 (35 min MUSE, 1 HST pointing in WFC3 F105W, F555W, F814W) and ACT-CL J0106.7+0103 (15 min MUSE, 1 HST pointing in WFPC2 F606W) \citep[all detected with ACTPol,][]{ACTPol_cl}. The MUSE data might be located at different places within the clusters depending on the original proposal (some target strong lensing, some nearby QSOs, and so on). 

\item  \textbf{DEEP2-2h} (PSTZ) \\ \votelss\votepz\votewl \\
Overlapping data: DEEP2, KiDS-VIKING, DES Deep Field, HSC \\
DEEP2 \citep{Newman:13} Fields 3 and 4 contain large, targeted spectroscopic datasets that are useful for photo-$z$ training.
Deemed a crucial redshift calibration field for KiDS-VIKING high-redshift galaxies \citep{Hildebrandt2019}, and selected as one of the deep, NIR-overlapping, SN fields for the Dark Energy Survey (Hartley, Choi \& Amon in prep.), this field is useful for both redshift and shear calibration comparisons. 

\item \textbf{DEEP2-23h} (Z) \\ \votelss\votepz\votewl \\ 
Overlapping data: DEEP2, PRIMUS, KiDS-VIKING, VIDEO, OzDES\\
DEEP2 \citep{Newman:13} Fields 3 and 4 contain large, targeted spectroscopic datasets that are useful for photo-$z$ training.
Deemed a crucial redshift calibration field for KiDS-VIKING with high-redshift galaxies \citep{Hildebrandt2019}, this field is useful for both redshift and shear calibration comparisons.

\item  \textbf{DES SN S1 / S2} (Z) \\ \votepz \\
Overlapping data: S82, OzDES, DES\\
DES supernova fields in the Stripe 82 region.

\item \textbf{DES 0408-5354} (I) \\ \votesl

DES 0408-5354 is both a multiply-imaged quasar and a compound lens. The quasar time delays are $>$ 40 days (Courbin et al. 2018). Observations of this target during commissioning would be useful for testing lens-detection algorithms, even if the cadence and time baseline are not sufficient to measure the time delay.

\item \textbf{Euclid Deep Field South} (SZ) \\ \votesn \\
Euclid Deep Field South (EDFS)\footnote{\url{https://www.cosmos.esa.int/web/euclid/euclid-survey}} is a DESC priority for a 5th Deep Drilling Field \citep[see][]{DESCobs, Jain, Rhodes, Capak}.
EDFS is the largest (23~deg$^2$) of three Euclid Deep Drilling fields selected by the Euclid Consortium for repeated, deep space-based optical and near-IR observations.

\item \textbf{Frontier Fields / CLASH / BUFFALO / RELICS clusters} (IS[Z]) \\ \votecl\votewl \\
\textbf{} \\
Overlapping data:  HST, DES Deep Field, VIDEO (NIR)\\ 
Clusters in these samples have been targeted with deep multi-band HST imaging, and have a wealth of other data.  A370, MACS~J0416.1-2403, A2744,  RXJ 2248-4431 (also known as AS1063) are part of the Hubble Frontier Fields \citep{HFF}, providing deep 7-band imaging in a single central pointing and a parallel field.  The BUFFALO survey expands these datasets by providing 2x2 HST mosaics for both central and parallel fields for each cluster, in 5 bands each.  MACS~J0416.1-2403 and RXJ 2248-4431 are part of the CLASH survey \citep{CLASH}, with HST imaging of the central field in 16 bands, and $\sim 5000$ / $\sim 3600$ VLT VIMOS spectroscopic redshifts \citep{rosati14}.  MACS~J0417.5-1154 has comparable HST imaging through the RELICS program \citep{RELICS}.  Additional datasets, such as multi-band Subaru imaging, extensive spectroscopy, and Spitzer imaging are available from these HST programs through the respective STScI websites.  In addition, all of these cluster fields have MUSE IFU spectroscopy available.  MACS~J0417.5-1154, A370, and MACS~J0416.1-2403 are early JWST targets.
    
\item \textbf{Galactic Bulge} (AT) \\ \votedkm \\ 
Overlapping data: MACHO, EROS, DECam PALS, DECaPS\\
The Galactic bulge provides a dense stellar field during the austral winter months (May - July) that can be observed at low airmass. This region provides tests of crowded field data processing algorithms and can be used for astrometric calibration tests. We desire repeated observations for a variety of cadences to test difference imaging. Overlap with MACHO \citep{Alcock:2001}, EROS \citep{2006A&A...454..185H}, the DECam Plane Survey \citep[DECaPS;][]{Schlafly:2018}, and the DECam Paralensing Survey (PALS)\footnote{\url{https://www.noao.edu/noaoprop/abstract.mpl?2018A-0273}}.
    
\item \textbf{HE 0230-2130} (I) \\ \votesl

HE 0230-2130 is a strongly lensed quasar with four images, separated by less than $\sim$2". One time delay between the multiple images was measured by Millon et al. (2020; arXiv:2002.05736) to be $\sim$16 days. Observations of this target during commissioning would be useful for testing lens-detection algorithms and also the building of quasar light curves to measure/verify the time delay.

\item \textbf{HSC J142449-005322} (I) \\ \votesl

HSC J142449-005322 is a compound lens system with two background galaxies at different redshifts being strongly lensed by the same foreground galaxy. The lensing arcs of two sources have radii of $\sim$1" and $\sim$3", respectively. Observations of this target during commissioning would be useful for testing lens-detection algorithms.

\item \textbf{HSC-Wide Fall-eq/VVDS/S82} (W) \\ \votelss\votepc\votepz\votesn\votewl \\
Overlapping data:   VVDS, HSC, DES, CFHTLeNS, CS82, SDSS, BOSS, WiggleZ \\
The Hyper Suprime-Cam SSP Wide Survey \citep{HSCDR1}  Fall region provides multi-band imaging from HSC over a wide area and contains spectroscopy from WiggleZ \citep{Drinkwater:10}. This field is useful for validation of  cross correlation-based photo-$z$ calibration techniques. For SN, observations of this field would test wide-field operations.
The wide area of this field, together with the rich spectroscopic sample available, will be useful for calibrating large-scale systematic effects, in comparison with the multiple existing weak-lensing and clustering measurements.

\item  \textbf{HSC-Wide Fall-eq/CFHT-W1/XMM} (W) \\
\votelss\votepz\votewl \\
Overlapping data:  XMM, UKIDSS, HSC, DES, CFHTLeNS, BOSS, WiggleZ \\
As above.

\item  \textbf{G9 (KIDS+VIKING+HSC+CFHTLenS)} (W) \\
\votelss\votepz\votewl\\\
Overlapping data:  HSC, Herschel, KiDS-VIKING, CFHTLenS, SDSS, BOSS, GAMA-09 \\
While the Spring region overlaps with spectroscopic redshift from GAMA \citep{Liske:15}, both regions are useful for validation of  cross correlation-based photo-$z$ calibration techniques.

\item\textbf{KiDES2DF} (W) \\ \votelss\votewl \\
Overlapping data:  KIDS, DES, 2dFLenS \\
Overlapping imaging from DES and KiDS-VIKING, as well as spectroscopy from 2dFLenS \citep{2dFLenS} are useful for validation and cross correlation over a wide area.

\item \textbf{LMC} (AT)\\ \votedkm \\
Overlapping data: MACHO, EROS \citep{2007A&A...469..387T}, SMASH, DECam PALS.  \\
The LMC provides a dense stellar field available during the austral summer months (October - February), which facilitates microlensing studies. Due to the near-polar location of the LMC, these fields will necessarily be observed at high airmass. We would desire repeated observations over a variety of cadences to test difference imaging, as well as shorter exposure times (compared to the  observations of the MACHO group by the DECam PALS) in order to access variability on a broader range of timescales (corresponding to a range of MACHO masses).

\item \textbf{LSST DDF CDF-S/GOODS-S/E-CDF-S} (ISTZ) \\ \votelss\votepc\votepz\votesn\votewl \\ 
Overlapping data: HST-GOODS-S, DES-Deep Field, VIDEO, OzDES, HST-CANDELS, VANDELS, MUSE-Wide\\
GOODS-S \citep{Dickinson:03} contains multi-band HST imaging (including HST-GEMS in F814W), multiwavelength ground-based imaging, and spectroscopic redshifts from VVDS-wide, FORS2, and other surveys to $i_{AB}\sim 24.0$ for photo-$z$ training.
It also has the deepest imaging available, from the HUDF \citep{Beckwith:06}, including MUSE GTO data. This field is interesting for accurate calibration of the deblending pipeline and artifact rejection.

\item \textbf{LSST DDF COSMOS} (PSTZ) \\ \votelss\votepc\votepz\votesn\votewl \\ 
Overlapping data: HST-CANDELS, zCosmos, C3R2, UltraVista, VVDS, PRIMUS, Spitzer, KiDS-VIKING, DES-Deep Field, HSC, PS1 MD04, MUSE-Wide\\
This field is extremely rich in both deep multiwavelength imaging (including extensive HST coverage) and spectroscopy. The COSMOS 30-band photometric redshifts \citep{Laigle:16} are extensively relied upon by multiple stage III lensing surveys for photometric redshift training, which makes this field particularly valuable both for photo-$z$ training and comparison with previous surveys.  
HST imaging allows for detailed deblending analyses to understand the impact of this systematic on WL. In addition, lensing signals can be compared directly with determinations from HSC \citep{HSCDR1} using similar pipelines, as well as DES \citep{DESDR1} and KiDS \citep{KiDSDR4} to varying depths. The extensive multiwavelength coverage enables detailed study of galaxy SEDs.

\item \textbf{LSST DDF ELAIS S1} (TZ) \\ \votepc\votepz\votesn\votewl\\
Overlapping data:  UKIDSS-DXS (NIR), LOFAR-Deep (radio), DES Deep Field, VIDEO \\

\item\textbf{ LSST DDF XMM-LSS / VVDS-Deep / VIPERS } (TZ)  \\ \votepz\votewl\votelss\votesn\votepc\\ 
Overlapping data: XMM, VIPERS, VVDS, PRIMUS, UKIDSS-DXS, DES-Deep Field, VIDEO (NIR), OzDES, SNLS D1, Pan-STARRS MD01, C3R2\\
The VIMOS VLT Deep Survey \citep[VVDS][]{LeFevre:13} has nearly 10,000 secure redshifts for galaxies to $i_{AB}\leq 24.75$ that would be useful for photo-$z$ training.\\
Spectra obtained by VIPERS in this field for more than 90k galaxies (red magnitude I(AB) $ < 22.5$) over $\sim$24~deg$^2$ facilitates redshift calibration. Nine-band WL data from KiDS-VIKING and 8-band from DES-VIDEO-Deep Field allow for shear calibration comparisons. C3R2 includes a very deep ($i<24.5$) survey designed to cater to weak lensing needs for an unbiased, deep spectroscopic catalog for photo-$z$s.

\item  \textbf{MS0451-03} (ISZ) \\ \votecl \\
Overlapping data:  HST , DEIMOS (spec), Magellan/IMACS (prism spec) \\ 
This cluster at $z=0.538$ has been imaged with a large, contiguous HST mosaic consisting of 42 HST/ACS pointings in F814W, extending out to 5~Mpc from the cluster center \citep{Moran07}, which allows testing deblenders and shear estimates.  It also has been targeted with DEIMOS spectroscopy \citep[][]{Moran07,Crawford11} ($\sim1400$ redshifts, and Magellan/IMACS prism spectroscopy ($\sim 2000$ redshifts, Herbonnet et al., in prep.), useful for validating redshift analyses.  In addition, it has deep, weak-lensing quality SuprimeCam imaging in BVRIz \citep{WtG1} as well as multiwavelength coverage from GALEX NUV \citep{Moran07} to VIRCAM (Nonino et al. in prep.) and Spitzer \citep{Wardlow10}.  It thus spans a similar range in wavelength as the COSMOS dataset, which could provide similar high-quality photometric redshifts.

\item \textbf{PS 1606-2333} (I) \\ \votesl

PS 1606-2333 is a strongly lensed quasar with four images separated by less than $\sim$2". The three time delays between the multiple images are all measured by Millon et al. (in prep) and span between $\sim$10--45 days. Observations of this target during commissioning would be useful for testing lens-detection algorithms and also building quasar light curves to measure/verify the time delay(s).

\item  \textbf{RX J0152.7-1357} (IS)\\\votecl \\
    Overlapping Data: HST, Spitzer, Herschel \\ This cluster at $z\sim 0.8$ has been imaged with a two-band HST mosaic over a $\sim 10^{\arcmin}$ field, multi-band HST coverage of the core, as well as deep X-ray (Chandra and XMM), Spitzer, and Herschel imaging. It has $>2000$ LDP redshifts to $i<23.75$ \citep{patel09}, as well as several hundreds of redshifts from VLT, Gemini and others. Deep SuprimeCam imaging. Also known as WARP J0152.7-1357 and CLJ01527-1357.  This massive cluster at high-z, will provide a good test of our ability to detect and separate cluster galaxies when the 4000~\AA break is in i-band.  It has known multiply imaged strong lensing systems.

\item \textbf{SDSS J0924+0219} (I) \\ \votesl

SDSS J0924+0219 is a strongly lensed quasar with four images separated by less than $\sim$2". One time delay between the multiple images was measured by Millon et al. (2020; arXiv:2002.05736) to be $\sim$2 days. Observations of this target during commissioning would be useful for testing lens-detection algorithms and also building quasar light curves to verify the time delay.


\item \textbf{SMC} (AT)\\ \votedkm \\
    Overlapping data: MACHO, EROS, SMASH, DECam PALS.  \\
    The SMC has intermediate stellar density and is well matched in angular size to the LSST focal plane. The SMC is observable during austral summer at high airmass. We desire repeated observations over a variety of exposure
    times and cadences. See LMC for details.

    \item \textbf{SPT clusters} (ISZ) \\ \votecl \\
    Overlapping Data: HST, DES, VLT, SPT, ACT, Chandra, spectroscopic observations, some Herschel/SPIRE, Spitzer/IRAC [I1],[I2] \\ All have X-ray temperatures ($\gtrsim$2k counts/cluster), spectroscopic redshifts, 4 pointings in F606W, F814W from HST/ACS ($\sim$2~ks/filter), VLT/FORS2 I$_\textrm{BESS}$. These are 13 of the most massive high-$z$ clusters selected by the SPT, with $z_\textrm{median}=0.88$ \citep[0.576 $<z<$1.132; see][]{2018MNRAS.474.2635S}.

\item \textbf{Stripe 82 / early DESI fields} (Z) \\ \votepz \\ Having wide area data with spectroscopic redshifts from DESI will be key to early tests of the cross correlation-based redshift calibration techniques that are essential for DESC science.  


\item  \textbf{SXDS / UKIDSS UDS / C3R2} (ISZ) \\ \votelss\votepz\votewl\\ 
Overlapping data: UKIDSS-UDS (NIR), SpUDS (MIR), VVDS (spec), PRIMUS (spec), UDSz (spec) HST-CANDELS, VANDELS, DES \\
The Subaru-XMM-Newton Deep Field \citep[SXDS or SXDF][]{Furusawa:08} contains HST imaging from CANDELS \citep{Grogin:11}, extensive multiwavelength imaging, and spectroscopy from UDSz \citep{uds}, VANDELS, and other surveys.  This is a key spectroscopic training field, and is also the location of DES SN Field X3. The C3R2 survey includes very deep spectroscopy which will be very important for photo-$z$ training and testing. N.B. This field is only 1.22 degrees from LSST DDF XMM-LSS, and so should be within the same FOV for LSSTCam, but not ComCam. 
This field also contains the cluster target SL2SJ021737-051329.

\item \textbf{WFI 2026-4536} (I) \\ \votesl

WFI 2026-4536 is a strongly lensed quasar with four images, separated by less than $\sim$2". One time delay between the multiple images was measured by Millon et al. (2020; arXiv:2002.05736) to be $\sim$19 days. Observations of this target during commissioning would be useful for testing lens-detection algorithms and also building quasar light curves to measure/verify the time delay.

\end{itemize}

\section{Accompanying Material}
\label{sec:Appendix-Accompanying-Material}

The LaTeX source of this DESC note along with the spreadsheets used to collect suggested fields from DESC members, and the Jupyter notebooks used to generate the tables and plots are available to the reader. The LaTeX source is available in the arXiv posting, and the associated files are available on Zenodo at the following DOI: \url{doi:10.5281/zenodo.4148042}.

In the Zenodo repository you will find: 

\begin{enumerate}
    \item A PDF version of the note.
    \item An exported Google sheet containing the suggested fields with the name 'Commissioning targets export version 10 20.xlsx'.
    \item Two Jupyter notebooks: airmass\_calendar\_noURL.ipynb (used to make the tables and the *.csv files), and visibility\_figures.ipynb used to make the observability figures.
    \item A copy of the csv files generated by the airmass notebook (and used by the visibility notebook to make the figures).
\end{enumerate}

To use the exported Google sheet, you can open it in Excel, but you will only be able to view it. If you would like it to be a live document, modify it, and access the information via the notebooks, you should: from your Google Drive select "New" and then "File Upload".  After it is opened, select "File $->$ Save as google sheets".  You should now have a live version of the spreadsheet you can access via the Jupyter notebooks. 

\subsection{Description of the spreadsheet}

In the spreadsheet, you will find several tabs.  Each DESC working group entered their suggested fields along with accompanying backup information, references and rationale.  Every tab has a minimum number of columns in common with all other tabs, so that they can be merged. But any given tab may also contain extra columns.

Descriptions of the tabs are as follows:

\begin{itemize}
    \item All - All of the entries from the other tabs.  There is one line per entry and fields that are requested by multiple groups are duplicated here.
    \item Merged - All of duplicated requests are merged and the numbers of working groups that requested them are also calculated.
    \item Notes - Useful supplemental information
    \item Template -  A blank template for groups to work from
    \item Calibration - Requests from the Photocalibration group
    \item CL - Requests from the Clusters group
    \item CL Lower Priority - Not used in the Note, but contains supplemental cluster fields
    \item DKM - Request from the Dark Matter group
    \item LSS - Requests from the Large Scale Structure group
    \item PZ - Requests from the Photo-Z group
    \item SL - Requests from the Strong Lensing group
    \item SN - Requests from the Supernova group
    \item WL - Requests from the Weak Lensing group
    \item Priority - Used to generate the sky map of priority fields
\end{itemize}

\subsection{Using the Notebooks}

The notebooks can read the tabs of the spreadsheet directly from the web. If you would like to read from your own copy then you should select ``File $->$ Publish to the web" Choose the tab you would like a link to and then select ``Comma-separated values (.csv)" as an option. For example, in the notebook airmass\_calculator.ipynb, modify the URLs in the ``Get Target Fields" cell to point at your cells.  The ``Merged Sheet" is the tab used for calculations in this note.  The airmass\_calendar.ipynb notebook includes several example visualizations to help plan observations, and can optionally output .csv files with information on the visibility of targets throughout the calendar year (the observability.csv and observability\_spectrophotometric.csv files were both created with this notebook). We also used the airmass\_calculator.ipynb notebook to generate the contents of the candidate target tables in this note.

\end{document}